\documentclass[aps,pra,twocolumn,superscriptaddress]{revtex4-2}

\usepackage{graphicx}
\usepackage{amssymb}
\usepackage{amsmath}
\usepackage{multirow}
\usepackage{braket}

\usepackage{array}
\newcolumntype{P}[1]{>{\centering\arraybackslash}p{#1}}

\begin{document}

\title{Wave packet dynamics and long-range tunneling within the SSH model using Rydberg-atom synthetic dimensions}
\author{Y. Lu}
\author{C. Wang}
\author{S. K. Kanungo}
\affiliation{Department of Physics and Astronomy, Rice University, Houston, TX  77005-1892, USA}
\author{S. Yoshida}
\affiliation{Institute for Theoretical Physics, Vienna University of Technology, Vienna, Austria, EU}
\author{F. B. Dunning}
\author{T.~C.~Killian}
\affiliation{Department of Physics and Astronomy, Rice University, Houston, TX  77005-1892, USA}

\begin{abstract}
Rydberg-atom synthetic dimensions in the form of a lattice of n$^3S_1$ levels, $58\leq n \leq 63$, coupled through two-photon microwave excitation are used to examine dynamics within the single-particle Su-Schrieffer-Heeger (SSH) Hamiltonian.  This paradigmatic model of topological matter describes a particle hopping on a one-dimensional lattice with staggered hopping rates.  Tunneling rates between lattice sites and on-site potentials are set by the microwave amplitudes and detuning, respectively.  An atom is first excited to a Rydberg state that lies within the lattice and then subject to the microwave dressing fields.  After some time, the dressing fields are turned off and the evolution of the population distribution in the different final lattice sites monitored using field ionization.  The measurements show the existence of long-lived symmetry-protected edge states and reveal the existence of direct long-distance tunneling between the edge states.  The results are in good agreement with model calculations and further demonstrate the potential of Rydberg-atom synthetic dimensions to simulate and faithfully reproduce complex Hamiltonians.
\end{abstract}

\maketitle

\section{Introduction}
\label{S:into}
Interest in the simulation of condensed matter behavior using atomic, molecular, and optical systems has increased dramatically in recent years.  Many of these studies utilize synthetic dimensions that are based on discrete states in an internal or external degree of freedom within a system that are coupled together and reinterpreted as lattice sites along an artificial spatial dimension.  Studies show that such synthetic dimensions can provide a powerful tool to examine the  behavior of many different physical systems.

Synthetic dimensions have been realized using a variety of platforms based on cold atoms and molecules, as well as photonic systems \cite{bcl12,opr19}.  In the case of cold atoms, for example, synthetic dimensions have been realized experimentally using hyperfine \cite{bcl12,mpc15,sla15}, clock  \cite{lcd16,kbb17}, magnetic \cite{cse20}, and Rydberg states \cite{kwl22,yrb12}, as well as with trap \cite{ose23}, momentum \cite{mag16,vsc19}, orbital \cite{khs20}, and superradiant states \cite{clw19}.  These have been used to explore a number of exotic physical phenomena, including quantum Hall ladders \cite{mpc15,sla15,mag16}, topological Anderson insulators \cite{mei18}, and higher-dimensional lattice physics \cite{bcl12,vsc19,pzo18,chv23}.

Recently a synthetic dimension was demonstrated that utilized a lattice of microwave-coupled $^3S_1$ and $^3P_1$ Rydberg states created within a single $^{84}$Sr atom \cite{kwl22}.  
In the Rydberg atom realization the tunneling rates were controlled by varying the microwave power and the on-site potentials set by the microwave detuning.  Alternating strong and weak coupling between the levels was used to simulate the single-particle Su-Schrieffer-Heeger (SSH) Hamiltonian which describes hopping on a one-dimensional lattice with staggered hopping rates and forms a paradigmatic model of topological matter \cite{ssh79}.  In the case of weak tunneling to the edge states, the  model possesses edge states that have a topologically-protected degeneracy in the limit of a large lattice and that are robust against perturbations respecting the chiral symmetry of the tunneling pattern \cite{ssh79,cts16}.  Such behavior has been observed in multiple systems \cite{kwl22,mag16,aab13,sgg17,llp19,hga23a}.  The first demonstration of particle interactions in Rydberg synthetic dimensions was recently reported \cite{chv23}.

Here we demonstrate that Rydberg atom synthetic dimensions can also provide  a powerful probe of particle dynamics.  As in our earlier work, we principally study a lattice of six coupled Rydberg states.  However, in contrast to these earlier studies, these now comprise a series of $^3S_1$ states that are coupled by two-photon microwave transitions.  As will be shown, use of $^3S_1$ states allows the population in each individual level, i.e., synthetic lattice site, to be separately identified through selective field ionization.  (In the earlier work the $(n+1)^3S$ and $n^3P$ states could not be separately resolved.)  In the present work individual Rydberg $n^3S_1$ states with a value of $n$ that corresponds to one of the lattice sites are initially excited and then subject to microwave dressing.  The subsequent evolution of the distribution of excited states within the coupled manifold is measured.  The measurements reveal the distinct dynamics of bulk and topologically-protected edge states.  Striking long-range tunneling is observed between the edge states.  This probes the energy-splitting and coupling between the two states, which decrease with system size, providing information on the localization of these states, which is of interest for their use as quantum information resources.  Destruction of their topological protection is demonstrated when the chiral symmetry is broken by detuning the microwave fields from resonance.

\section{Experimental method}
\label{S:experimental}
The present apparatus, 
\begin{figure}
\includegraphics[width=8.5cm]{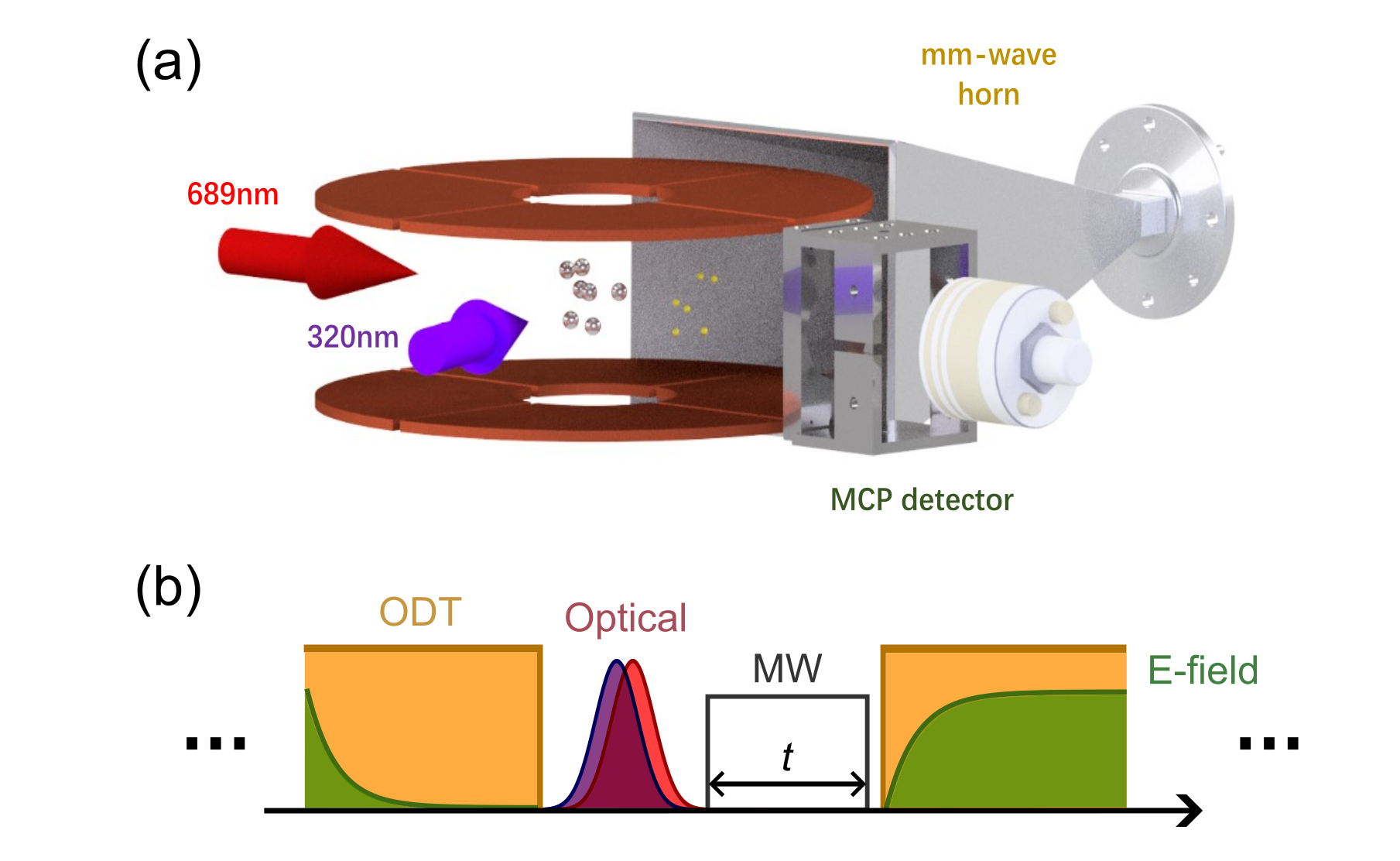}
\caption{
\label{fig:schematic}
a) Schematic of the experimental apparatus.  b) The time sequence of one experimental cycle. The ODT is turned off during Rydberg excitation and micro-wave (MW) application. 
The electric field plates are fully discharged before the start of the next experimental cycle.}
\end{figure}
aspects of which are described in more detail elsewhere \cite{ssk13,emm09,dwk18}, is shown in Fig.~\ref{fig:schematic}.  Briefly, $^{84}$Sr atoms emerging from a Zeeman slower are cooled in two stages, first using a ``blue'' magneto-optical trap (MOT) operating on the 461~nm 5s$^2$ $^1S_0$ - 5s5p $^1P_1$ transition, followed by cooling in a narrow-line ``red'' MOT operating on the 689~nm 5s$^2$ $^1S_0$ - 5s5p $^3P_1$ transition.  
The atoms are then loaded into a 1064~nm crossed-sheet ``pancake'' optical dipole trap.  The final atom temperature is controlled by evaporative cooling yielding cold atom clouds with $\sim10^5$ atoms,
peak densities $\sim 10^{11}$~cm$^{-3}$, and temperature T $\sim1~\mu$K. 

The Rydberg levels forming the lattice are coupled by microwave-driven two-photon transitions.  The required microwave frequencies are generated by combining the outputs of a series of RF synthesizers.  The atoms are irradiated using a K-band horn antenna.  The accessible frequency range is $\sim13-26$~GHz.  Each coupling strength and resonant frequency is calibrated using Autler-Townes splitting in a two-level configuration \cite{kwl22}. However, the addition of the other frequencies leads to AC Stark shifts that are determined experimentally and compensated through an iterative process.

A magnetic field is applied to allow selective excitation of individual magnetic sublevels.  The applied field, $\sim6$~Gauss, shifts the sublevels by $\sim{17}$~MHz, which is large compared to the tunneling rates.  The microwave frequencies typically are adjusted to maintain resonant couplings but, if required, detuning can be purposely introduced.

\begin{figure}
\includegraphics[width=8cm]{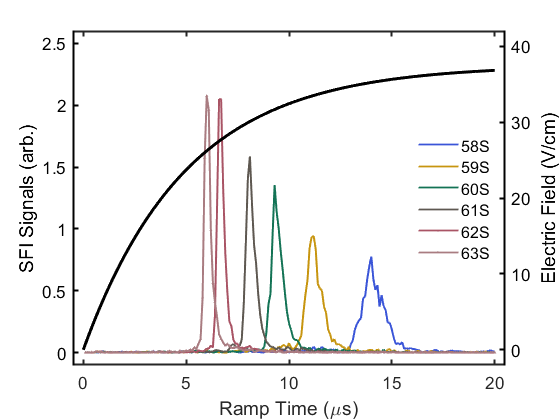}
\caption{
\label{fig:SFI spectra} SFI spectra recorded using the individual states of interest in the present work.  The figure also shows the time evolution of the applied field.}
\end{figure}

The initial n$^3S_1$ Rydberg states are created by two-photon excitation via the intermediate 5s5p$^3P_1$ state (with a blue detuning of 80~MHz) and laser polarizations chosen to create $n^3S_1$
($m=+1$) states.  The Rydberg atoms are then subject to the microwave fields for a selected time $t$, following which they are detected through ionization in a ramped electric field of the form E($t$) = E$_p (1-e^{-t/\tau})$ 
where $\tau \sim 5~\mu$s 
and E$_p\simeq 40$~V~cm$^{-1}$ (see Fig.~\ref{fig:SFI spectra}).  
A $^3S_1$ Rydberg atom with quantum number $n$ will ionize adiabatically at a field given by $1/16(n-\delta)^4$~a.u. where $\delta$ is the quantum defect, $\delta(n^3S_1)  = 3.371$. The product electrons are directed to a microchannel plate for detection.  
Their arrival time defines the field at which ionization occurred and hence the initial state, i.e., lattice site.  A montage of the SFI spectra recorded using the individual states of interest in the present work is shown in Fig.~\ref{fig:SFI spectra}.  
These states are well-resolved, and by fitting a linear combination of these spectra to a given data set, the population distribution can be reliably determined and used to  monitor the evolution of the population in each individual state within the lattice.

Approximately $10^3$ measurement cycles, shown in Fig.\ref{fig:schematic}(b), can be performed using each cold atom sample at a repetition frequency of $\sim5$~kHz.  Fewer than one Rydberg atom, on average, is created per cycle to eliminate possible effects associated with Rydberg-Rydberg interactions.

\section{Model Calculations}
\label{s:Model Calculations}
\begin{figure*}
\includegraphics[width=12cm]{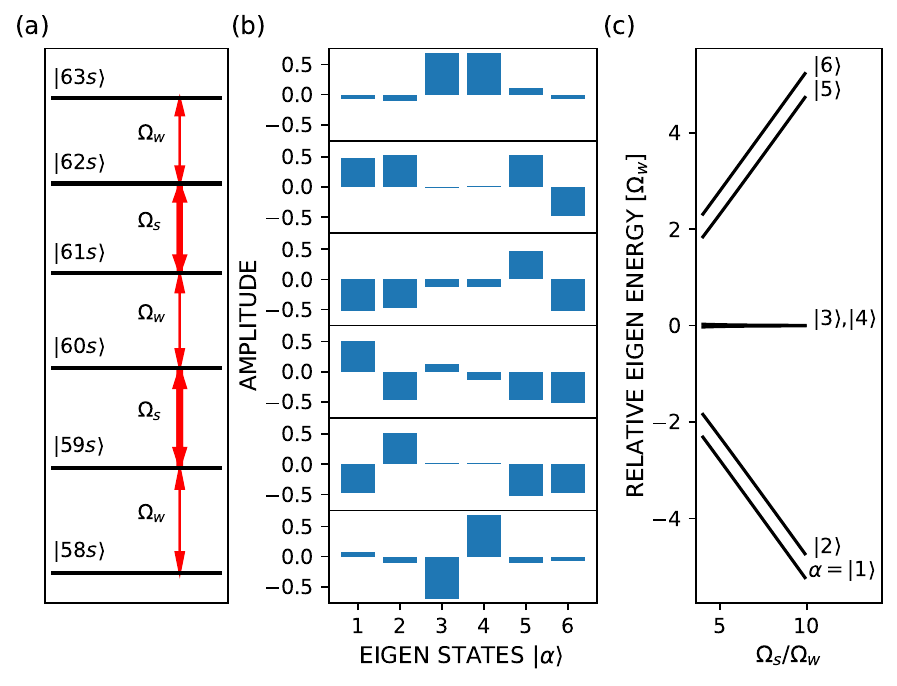}
\caption{(a)  The bare six-level lattice used in the present study.  The coupling strengths are $\Omega_W = 160$~kHz, $\Omega_S = 800$~kHz.  (b) Projection of the initial individual $\vert ns\rangle$ bare states into the dressed state basis, $\vert1\rangle$ to $\vert 6\rangle$.  The bars show the overlap amplitudes $\langle\alpha|ns\rangle$ (see Eq.~\ref{eq:fields on}), which may be chosen to all be real, and their phases relative to each other. (c) Energy level structure in the dressed-state basis expressed as a function of $\Omega_S/\Omega_W$
\label{fig:bare}
}
\end{figure*}

The SSH model consists of a series of states of arbitrary number connected by alternating weak and strong couplings as shown in Fig.~\ref{fig:bare}   
and can be considered as a chain of $N$ unit cells, each unit cell hosting two sites that lie on different sub-lattices \cite{aop16} which we define as the even and odd numbered sites.  In the bare state basis, and in the appropriate rotating frame, the general Hamiltonian describing the system may be written
\begin{equation}
    \hat{H} = \sum^{2N-1}_{n=1}\frac{\Omega_{n,n+1}}{2}\left(\hat{c}_{n}^{\dagger}\hat{c}_{n+1} + h.c. \right) + \sum^{2N}_{n=1}\hat{c}_{n}^{\dagger}\hat{c}_{n}U_{n} 
    \label{eq1}
\end{equation}
where $2N$ is the total number of bare states, $\Omega$ is the Rabi frequency (coupling)  and $U$ is the on-site interaction (local potential) term.  (The Rabi frequency is related to the tunneling rate by $\Omega=-2J$.) In the present case of a six-site ladder, Eq.(1) may be written in matrix form, resulting in the Hamiltonian matrix:
\begin{equation}
    \langle\hat{H}\rangle =
    \begin{bmatrix}
    U_{58s} & \Omega_{W}/2 & 0 & 0 & 0 & 0 \\
    \Omega_{W}/2 & U_{59s} & \Omega_{S}/2 & 0 & 0 & 0 \\
    0 & \Omega_{S}/2 & U_{60s} & \Omega_{W}/2 & 0 & 0 \\
    0 & 0 & \Omega_{W}/2 & U_{61s} & \Omega_{S}/2 & 0 \\
    0 & 0 & 0 &\Omega_{S}/2 & U_{62s} & \Omega_{W}/2 \\
    0 & 0 & 0 & 0 & \Omega_{W}/2 & U_{63s}
    \end{bmatrix}
    \label{eq2}
\end{equation}
where $\Omega_W$ is the weak coupling Rabi frequency, $\Omega_S$ is the strong Rabi frequency, and the $U_n$ are the different on-site potentials that result from detuning of the different transitions.  If $U_{58s}$ is set to zero, then $U_{59s}$ corresponds to the detuning of the 58s-59s transition, etc.  In the majority of the present work the transition frequencies are carefully matched such that the $U_n$ are essentially zero, whereupon there is no connection within each of the two sub-lattices mentioned previously and chiral symmetry prevails.  We do, however, introduce non-zero values of the $U_n$ to examine their effects on tunneling rates and topological edge-state protection.
\begin{figure}
\includegraphics[width=8cm]{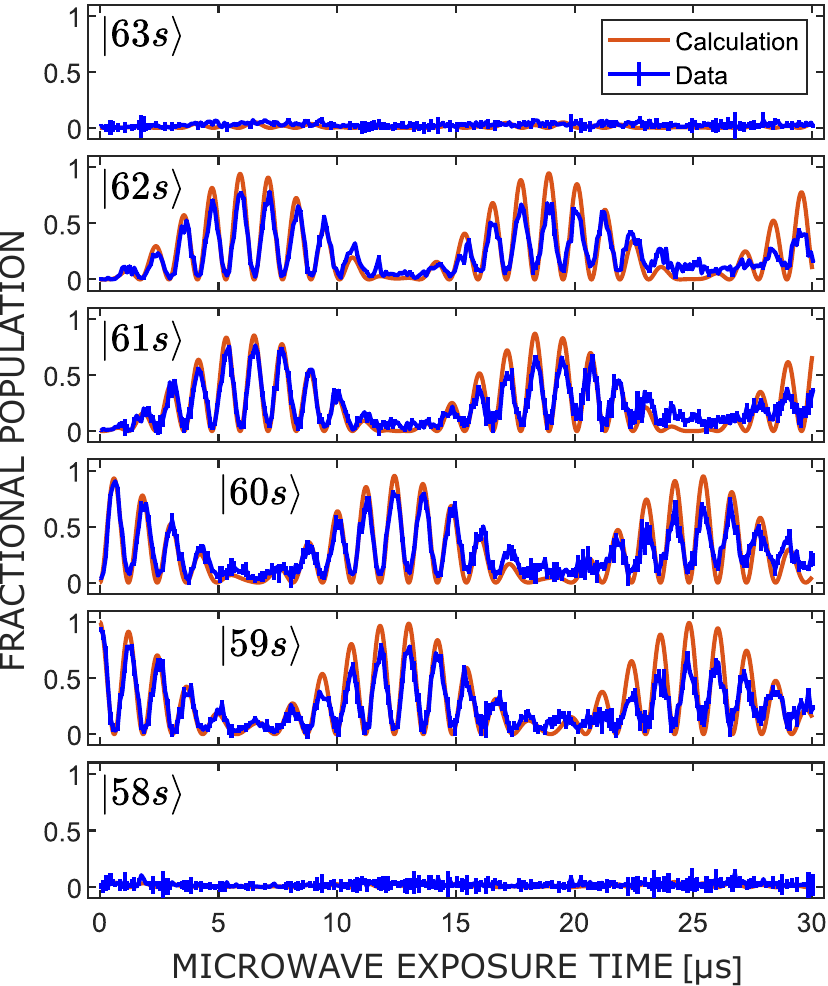}
\caption{
\label{fig:time dependence}
The fractional final populations (see text) in each bare lattice site (Rydberg state) following initial excitation of the $\vert59$s$\rangle$ state as a function of the microwave exposure time.  
}
\end{figure}
 The values selected, $\Omega_W\sim160$~kHz and $\Omega_S\sim 800$~kHz, provide a 5:1 ratio of strong-to-weak coupling strengths. 

The ``bare'' 5sns$^3$S$_1$ Rydberg states $\vert ns\rangle$, making up the present ladder are shown in Fig.~\ref{fig:bare}(a).  The two extreme bare states are termed ``edge states,'' those in the middle  ``bulk states.''  Initially a selected bare state is created.  At $t=0$ the microwave fields are turned on.  The dynamics can be understood by projecting an initial bare state $\vert ns\rangle$ into the dressed basis, i.e.,
\begin{equation}
\label{eq:fields on}
\vert ns\rangle = \sum_{\alpha=1}^{6} \vert\alpha\rangle\langle \alpha\vert ns\rangle
\end{equation}
where $\vert\alpha\rangle$ are the eigenstates in the presence of tunneling, which are labeled 1 to 6.  The amplitudes $\langle ns|\alpha\rangle $ and their relative phases are shown in Fig.~\ref{fig:bare}(b).  The structure of the dressed energy levels is shown in Fig.~\ref{fig:bare}(c) expressed as a function of $\Omega_S/\Omega_W$.  Given the present weak tunneling to the edge states, the SSH model predicts the presence of (quasi-)doubly-degenerate topologically-protected states $\vert 3\rangle$ and $\vert 4\rangle$ centered in the gap between the states $\vert 1\rangle$ and $\vert 2\rangle$, and $\vert 5\rangle$ and $\vert 6\rangle$ whose energies shift in direct proportion to $\Omega_S$.  
As shown in Fig.~\ref{fig:bare}, the $\vert58s\rangle$ and $\vert63s\rangle$ edge states principally project into the dressed states $\vert3\rangle$ and $\vert4\rangle$, whereas the $\vert59s\rangle$ to $\vert62s\rangle$ bulk states project into a mix of the $\vert1\rangle$, $\vert2\rangle$, $\vert5\rangle$, and $\vert6\rangle$ dressed states.  

After some microwave exposure time, $t$, the microwaves are extinguished and the superposition of dressed eigenstates is projected back into the  bare state basis, the projection into a given bare state $\vert ns^\prime\rangle$ being given by
\begin{equation}
    \label{eq:projection}
    \vert\langle ns^\prime \vert\hat{T}\vert ns\rangle\vert^2
    =\sum_{\alpha,\beta=1}^6 \langle ns \vert\alpha\rangle\langle\alpha\vert ns^\prime\rangle\langle ns^\prime\vert\beta\rangle\langle\beta\vert ns\rangle e^{+i\omega_{\alpha}t}e^{-i\omega_{\beta}t}
\end{equation}
where the $\vert\beta\rangle$ are the same eigenstates as $\vert\alpha\rangle$, $\hat{T}$ is the time evolution operator, and $\hbar\omega$ is the eigenenergy.  The energy separations between the eigenstates $\alpha,\beta = 1~\text{and}~2$, $\alpha,\beta = 3~\text{and}~4$,  and $\alpha,\beta = 5~\text{and}~6$ are small compared to those between $\alpha,\beta = 1~\text{or}~2~\text{and}~3~\text{or}~4$ and $\alpha,\beta = 3~\text{or}~4~\text{and}~5~\text{or}~6$. As a result, fast dynamics will be governed principally by the separation between the three ``groups'' of eigenstates, slow dynamics by the energy separations within a group.


\section{Results and Discussion}
\label{S:results}

Consider initially the case where the $\vert 59s\rangle$ Rydberg state is excited.  Figure~\ref{fig:time dependence} shows the time evolution of the relative populations in the different coupled bare states as a function of the time for which the microwave fields are applied.  The total Rydberg population, however, decays with a time constant of $\sim70 \mu$s predominently due to radiative decay. To compensate for this decay and better emphasize the population dynamics, Fig.~\ref{fig:time dependence} shows the evolution of the fractional populations in each $\vert ns\rangle$ level, i.e., at every point in time the total Rydberg signal is normalized to unity.

Inspection of Fig.~\ref{fig:time dependence} shows that, in essence, whereas at $t=0$ all the atoms are in the $\vert59s\rangle$ state, the strong coupling to the $\vert60s\rangle$ level results in rapid antiphase oscillations between their populations.  However, as time increases the weak coupling between the $\vert60s\rangle$ and $\vert61s\rangle$ states allows the wave packet to transfer into the strongly-coupled $\vert61s\rangle-\vert60s\rangle$ pair until, by $t\approx7\mu s$, essentially all the population density is entirely localized in this pair.  Further evolution leads to continuing periodic transfer of population between the $\vert59s\rangle-\vert60s\rangle$ and $\vert61s\rangle-\vert62s\rangle$ pairs.  Whereas strong oscillations are seen between the bulk states, no population transfer to the $\vert58s\rangle$ and $\vert63s\rangle$ edge states is evident.

As is seen in Fig.~\ref{fig:time dependence}, model calculations (Eq.~\ref{eq:projection}) reproduce well the observed behavior although at later times the amplitude of the population oscillations falls below that predicted by the model. This can be attributed to additional decoherence, a major source of which, ancillary calculations show, can be attributed to coupling to intermediate states during two-photon excitation.  Furthermore, radiative (or blackbody-radiation-driven) decay to nearby $^3$P states, provides an increasing background signal in the SFI spectrum, which again limits the amplitude of oscillations between the coupled states and the contrast in these oscillations. However, even at much later times ($t>70\mu s$), where there are essentially no visible oscillations between the bulk states, no significant population leaks into the edge states, demonstrating that topological protection remains robust even in the presence of decoherence. 

\begin{figure}
\includegraphics[width=8cm]{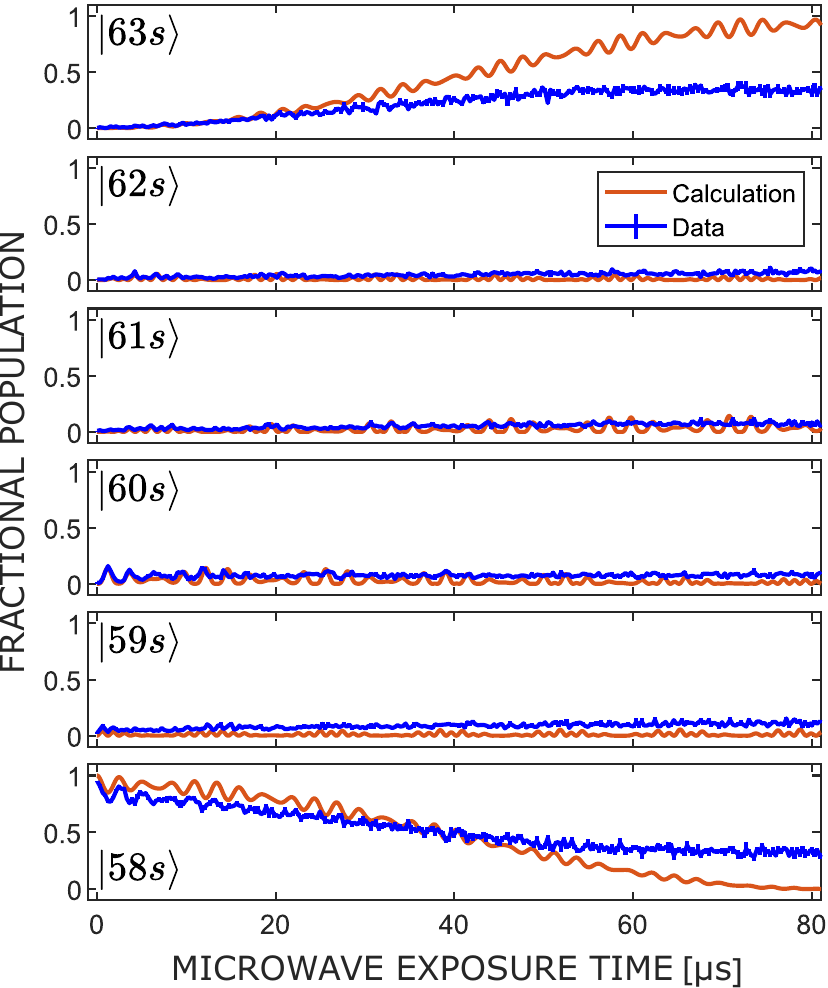}
 \caption{
\label{fig:behavior}
The final fractional population in each bare lattice site following initial preparation of the $\vert58s\rangle$ ``edge'' state as a function of microwave exposure time.  The model simulations assume that the coupling fields are on resonance.}  
\end{figure}

Figure~\ref{fig:behavior} shows the behavior observed 
following creation of the $\vert58s\rangle$ edge state.  Initially, for short microwave exposure times, the population remains largely localized in the edge state but begins to tunnel directly into the opposite edge state.  No significant population build-up in the intermediate bulk states is observed.  As shown in Fig.~\ref{fig:behavior}, such direct edge-to-edge population transfer is predicted by the simulations, the rate being determined by the energy separation between the (quasi-)degenerate dressed eigenstates $\vert3\rangle$ and $\vert4\rangle$.  Measurements showed that the resonance width of edge-to-edge tunneling with detuning of the 58s-59s coupling is extremely narrow, $\sim 5$~kHz full width at half maximum.  The lack of complete population transfer between the edge states at the later times can again be attributed to decoherence which results in a small build-up in the bulk state populations.

The small rapid (antiphase) oscillations in the $\vert58s\rangle$ and $\vert60s\rangle$ state populations are well reproduced by the calculations and result because, as shown in Fig.~\ref{fig:bare}, projection of the $\vert58s\rangle$ state into the dressed manifold leads to the small fractional population of the outermost dressed states, $\vert1\rangle$, $\vert2\rangle$, $\vert5\rangle$, and $\vert6\rangle$. The appearance of population only in the $\vert 60s \rangle$ bulk state on short timescales shows the emergence
of the characteristic exponentially-localized edge state with overlap on only one sublattice of the 
bipartite structure of the SSH model \cite{aop16}.

\begin{figure}
\includegraphics[width=8cm]{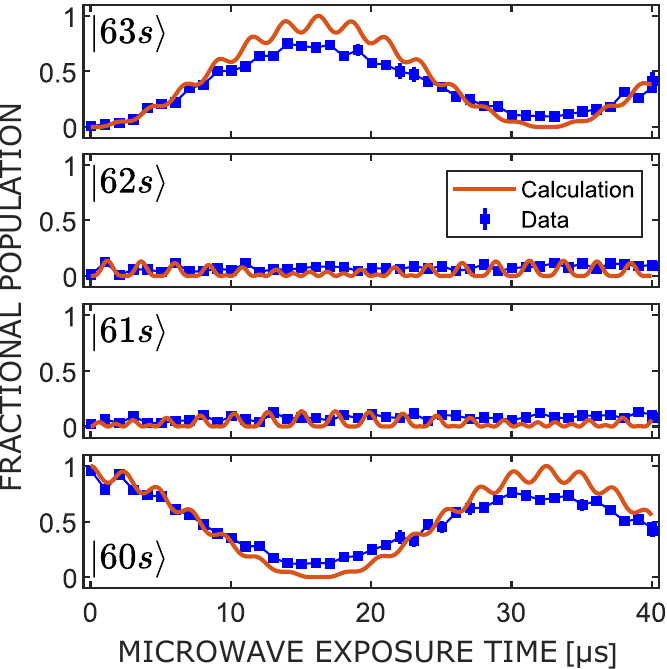}
\caption{
\label{fig:fractional}
The fractional populations in each bare lattice site following initial excitation of the $\vert 60s\rangle$ state using the four-level lattice $\vert60s\rangle-\vert63s\rangle$ and the coupling strengths shown in Fig.~\ref{fig:bare}.
}
\end{figure}

The edge-to-edge tunneling rate, i.e., the energy splitting between the dressed states $\vert3\rangle$ and $\vert4\rangle$, also depends sensitively on the number of states in the lattice.  Figure~\ref{fig:fractional}
shows the data recorded using a truncated four-level lattice that comprised only the $\vert60s\rangle-\vert63s\rangle$ states but with the same tunneling rates as in Fig.~\ref{fig:bare}.  Direct edge-to-edge tunneling is again seen and the results are in  excellent agreement with model predictions.  This is to be expected as the tunneling rates are higher and measurement times shorter than for the six-level lattice, limiting the effects of decoherence.  

\begin{figure}
\includegraphics[width=8cm]{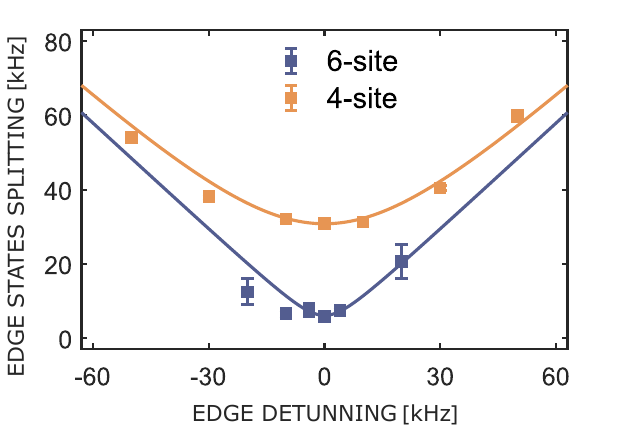}
\caption{
\label{fig:edge}
\textbf Energy splitting of the $\vert3\rangle$ and $\vert4\rangle$ edge states measured via the edge-to-edge tunneling rate as a function of the microwave frequency detuning from the $\vert58s\rangle - \vert 59s\rangle$ edge transition for the six-level lattice and $\vert60s\rangle-\vert61s\rangle$ transition for the four-level lattice (see text). The lines and data points show the theoretical calculations and experimental measurements, respectively.}
\end{figure}

Further experiments and model simulations were undertaken to explore the effects of edge-state detuning, and, through this, the robustness of edge-to-edge tunneling.  To this end, a series of measurements were performed in which the lower edge-state  transition was purposely detuned.  Tunneling rates measured as a function of this detuning are shown in  Fig.~\ref{fig:edge} for both the four- and six-level lattices.
In each case, edge-state detuning leads to an increased tunneling rate.  Furthermore, as the size of the lattice increases, the tunneling rate falls rapidly, decreasing by $\Omega_W /\Omega_S$ each time a pair of new lattice sites is added \cite{chv23}.  Indeed, in the limit that the number of lattice sites becomes very large, edge-to-edge tunneling will be suppressed.

\begin{figure}
\includegraphics[width=8cm]{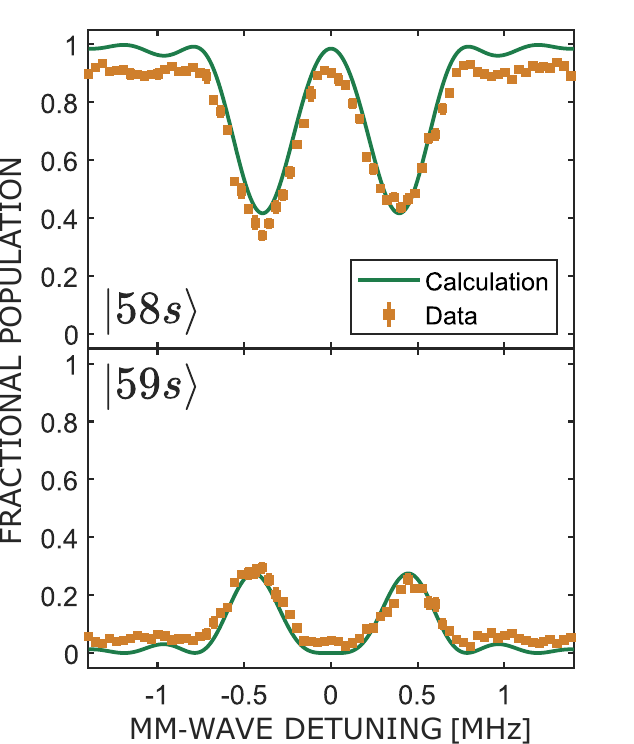}
\caption{
\label{fig:transfer}
Population transfer following initial population of the 58$^3S_1$ ``edge'' state as a function of the detuning of the $\vert58s\rangle - \vert 59s\rangle$ transition.  The other conditions are the same as for Fig.~\ref{fig:behavior}.  The populations were measured $2.5~\mu$s after turn-on of the microwave fields, at which time the majority of the population transfer is to the neighboring strongly-coupled $\vert59s\rangle$ and $\vert 60s\rangle$ states.  The solid lines show the results of model simulations.}
\end{figure}

With further detuning of the edge-state transition, chiral symmetry is broken and the edge states are no longer topologically protected, allowing population transfer from an edge state into the neighboring bulk states (and vice versa).  This behavior is demonstrated for the $\vert58s\rangle$ level in Fig.~\ref{fig:transfer}
which shows, for the six-level lattice, the final relative populations in the bulk and edge bare states, 
expressed as a function of the detuning, following application of a microwave field of 2.4~$\mu$s duration.  As is evident from the figure, a detuning of $\sim400$~kHz leads to rapid transfer of population from the $\vert58s\rangle$ state to the neighboring strongly-coupled $\vert59s\rangle$ and $\vert60s\rangle$ states.  (On the short time scale chosen, very little transfer to other lattice sites is seen.)  The measurements are again in reasonable agreement with model predictions.



\section{Conclusions}
\label{s:Conclusions}
The present work demonstrates the potential of Rydberg-atom synthetic dimensions as a tool with which to examine dynamics in quantum matter.  Topologically-protected edge states were observed in both a four- and six-site SSH model whose behaviors were consistent with that predicted by simulations.  Loss of this protection upon breaking of chiral symmetry was also observed.

The measurements lay the foundation for studies of more complex coupling schemes in which, for example, closed loops are generated by direct coupling of well-separated upper and lower lattice sites or ``impurities'' or domain walls are introduced by detuning one, or more, of the transitions or shifting the coupling pattern to have  adjacent weak couplings, giving rise to charge fractionalization \cite{jac07}.  It is also possible to create higher-dimensional synthetic lattices \cite{bcl12,tso10,jbu13} and to study systems with non-trivial spatial \cite{bcr15} and band structures \cite{bld16,qzh11} and higher-order topological states \cite{bhu17}.  The present platform is also suitable for the study of time-dependent phenomena such as Floquet symmetry-protected \cite{rha17} and non-equilibrium states \cite{mco19}, and can also be applied to the study of dynamics in interacting many-body systems \cite{sgh18,stw19} using ordered arrays of Rydberg atoms held in closely-spaced optical tweezers \cite{and22,bla20,kni21} or optical lattices \cite{ewl21,ssw21}.

 \begin{acknowledgments}
The authors are pleased to acknowledge valuable discussions with K. R. A. Hazzard during the course of this work which was supported by the NSF under Grant Nos. 1904294 and 2110596, the AFOSR under Grant No. FA9550-14-1-007, and the FWF (Austria) under Grant No. FWF-P35539-N, and Doctoral College FWF W 1243 (Solids4Fun).

\end{acknowledgments}

%


\begin{thebibliography}{43}%
\makeatletter
\providecommand \@ifxundefined [1]{%
 \@ifx{#1\undefined}
}%
\providecommand \@ifnum [1]{%
 \ifnum #1\expandafter \@firstoftwo
 \else \expandafter \@secondoftwo
 \fi
}%
\providecommand \@ifx [1]{%
 \ifx #1\expandafter \@firstoftwo
 \else \expandafter \@secondoftwo
 \fi
}%
\providecommand \natexlab [1]{#1}%
\providecommand \enquote  [1]{``#1''}%
\providecommand \bibnamefont  [1]{#1}%
\providecommand \bibfnamefont [1]{#1}%
\providecommand \citenamefont [1]{#1}%
\providecommand \href@noop [0]{\@secondoftwo}%
\providecommand \href [0]{\begingroup \@sanitize@url \@href}%
\providecommand \@href[1]{\@@startlink{#1}\@@href}%
\providecommand \@@href[1]{\endgroup#1\@@endlink}%
\providecommand \@sanitize@url [0]{\catcode `\\12\catcode `\$12\catcode
  `\&12\catcode `\#12\catcode `\^12\catcode `\_12\catcode `\%12\relax}%
\providecommand \@@startlink[1]{}%
\providecommand \@@endlink[0]{}%
\providecommand \url  [0]{\begingroup\@sanitize@url \@url }%
\providecommand \@url [1]{\endgroup\@href {#1}{\urlprefix }}%
\providecommand \urlprefix  [0]{URL }%
\providecommand \Eprint [0]{\href }%
\providecommand \doibase [0]{https://doi.org/}%
\providecommand \selectlanguage [0]{\@gobble}%
\providecommand \bibinfo  [0]{\@secondoftwo}%
\providecommand \bibfield  [0]{\@secondoftwo}%
\providecommand \translation [1]{[#1]}%
\providecommand \BibitemOpen [0]{}%
\providecommand \bibitemStop [0]{}%
\providecommand \bibitemNoStop [0]{.\EOS\space}%
\providecommand \EOS [0]{\spacefactor3000\relax}%
\providecommand \BibitemShut  [1]{\csname bibitem#1\endcsname}%
\let\auto@bib@innerbib\@empty
\bibitem [{\citenamefont {Boada}\ \emph {et~al.}(2012)\citenamefont {Boada},
  \citenamefont {Celi}, \citenamefont {Latorre},\ and\ \citenamefont
  {Lewenstein}}]{bcl12}%
  \BibitemOpen
  \bibfield  {author} {\bibinfo {author} {\bibfnamefont {O.}~\bibnamefont
  {Boada}}, \bibinfo {author} {\bibfnamefont {A.}~\bibnamefont {Celi}},
  \bibinfo {author} {\bibfnamefont {J.~I.}\ \bibnamefont {Latorre}},\ and\
  \bibinfo {author} {\bibfnamefont {M.}~\bibnamefont {Lewenstein}},\ }\href
  {https://doi.org/10.1103/PhysRevLett.108.133001} {\bibfield  {journal}
  {\bibinfo  {journal} {Phys. Rev. Lett.}\ }\textbf {\bibinfo {volume} {108}},\
  \bibinfo {pages} {133001} (\bibinfo {year} {2012})}\BibitemShut {NoStop}%
\bibitem [{\citenamefont {Ozawa}\ and\ \citenamefont {Price}(2019)}]{opr19}%
  \BibitemOpen
  \bibfield  {author} {\bibinfo {author} {\bibfnamefont {T.}~\bibnamefont
  {Ozawa}}\ and\ \bibinfo {author} {\bibfnamefont {H.~M.}\ \bibnamefont
  {Price}},\ }\href {https://doi.org/10.1038/s42254-019-0045-3} {\bibfield
  {journal} {\bibinfo  {journal} {Nat. Rev. Phys.}\ }\textbf {\bibinfo {volume}
  {1}},\ \bibinfo {pages} {349} (\bibinfo {year} {2019})}\BibitemShut {NoStop}%
\bibitem [{\citenamefont {Mancini}\ \emph {et~al.}(2015)\citenamefont {Mancini}
  \emph {et~al.}}]{mpc15}%
  \BibitemOpen
  \bibfield  {author} {\bibinfo {author} {\bibfnamefont {M.}~\bibnamefont
  {Mancini}} \emph {et~al.},\ }\href@noop {} {\bibfield  {journal} {\bibinfo
  {journal} {Science}\ }\textbf {\bibinfo {volume} {349}},\ \bibinfo {pages}
  {1510} (\bibinfo {year} {2015})}\BibitemShut {NoStop}%
\bibitem [{\citenamefont {Stuhl}\ \emph {et~al.}(2015)\citenamefont {Stuhl},
  \citenamefont {Lu}, \citenamefont {Aycock}, \citenamefont {Genkina},\ and\
  \citenamefont {Spielman}}]{sla15}%
  \BibitemOpen
  \bibfield  {author} {\bibinfo {author} {\bibfnamefont {B.~K.}\ \bibnamefont
  {Stuhl}}, \bibinfo {author} {\bibfnamefont {H.~L.}\ \bibnamefont {Lu}},
  \bibinfo {author} {\bibfnamefont {L.~M.}\ \bibnamefont {Aycock}}, \bibinfo
  {author} {\bibfnamefont {D.}~\bibnamefont {Genkina}},\ and\ \bibinfo {author}
  {\bibfnamefont {I.~B.}\ \bibnamefont {Spielman}},\ }\href@noop {} {\bibfield
  {journal} {\bibinfo  {journal} {Science}\ }\textbf {\bibinfo {volume}
  {349}},\ \bibinfo {pages} {1514} (\bibinfo {year} {2015})}\BibitemShut
  {NoStop}%
\bibitem [{\citenamefont {Livi}\ \emph {et~al.}(2016)\citenamefont {Livi},
  \citenamefont {Cappellini}, \citenamefont {Diem}, \citenamefont {Franchi},
  \citenamefont {Clivati}, \citenamefont {Frittelli}, \citenamefont {Levi},
  \citenamefont {Calonico}, \citenamefont {Catani}, \citenamefont {Inguscio},\
  and\ \citenamefont {Fallani}}]{lcd16}%
  \BibitemOpen
  \bibfield  {author} {\bibinfo {author} {\bibfnamefont {L.~F.}\ \bibnamefont
  {Livi}}, \bibinfo {author} {\bibfnamefont {G.}~\bibnamefont {Cappellini}},
  \bibinfo {author} {\bibfnamefont {M.}~\bibnamefont {Diem}}, \bibinfo {author}
  {\bibfnamefont {L.}~\bibnamefont {Franchi}}, \bibinfo {author} {\bibfnamefont
  {C.}~\bibnamefont {Clivati}}, \bibinfo {author} {\bibfnamefont
  {M.}~\bibnamefont {Frittelli}}, \bibinfo {author} {\bibfnamefont
  {F.}~\bibnamefont {Levi}}, \bibinfo {author} {\bibfnamefont {D.}~\bibnamefont
  {Calonico}}, \bibinfo {author} {\bibfnamefont {J.}~\bibnamefont {Catani}},
  \bibinfo {author} {\bibfnamefont {M.}~\bibnamefont {Inguscio}},\ and\
  \bibinfo {author} {\bibfnamefont {L.}~\bibnamefont {Fallani}},\ }\href
  {https://doi.org/10.1103/PhysRevLett.117.220401} {\bibfield  {journal}
  {\bibinfo  {journal} {Phys. Rev. Lett.}\ }\textbf {\bibinfo {volume} {117}},\
  \bibinfo {pages} {220401} (\bibinfo {year} {2016})}\BibitemShut {NoStop}%
\bibitem [{\citenamefont {Kolkowitz}\ \emph {et~al.}(2017)\citenamefont
  {Kolkowitz}, \citenamefont {Bromley}, \citenamefont {Bothwell}, \citenamefont
  {Wall}, \citenamefont {Marti}, \citenamefont {Koller}, \citenamefont {Zhang},
  \citenamefont {Rey},\ and\ \citenamefont {Ye}}]{kbb17}%
  \BibitemOpen
  \bibfield  {author} {\bibinfo {author} {\bibfnamefont {S.}~\bibnamefont
  {Kolkowitz}}, \bibinfo {author} {\bibfnamefont {S.~L.}\ \bibnamefont
  {Bromley}}, \bibinfo {author} {\bibfnamefont {T.}~\bibnamefont {Bothwell}},
  \bibinfo {author} {\bibfnamefont {M.~L.}\ \bibnamefont {Wall}}, \bibinfo
  {author} {\bibfnamefont {G.~E.}\ \bibnamefont {Marti}}, \bibinfo {author}
  {\bibfnamefont {A.~P.}\ \bibnamefont {Koller}}, \bibinfo {author}
  {\bibfnamefont {X.}~\bibnamefont {Zhang}}, \bibinfo {author} {\bibfnamefont
  {A.~M.}\ \bibnamefont {Rey}},\ and\ \bibinfo {author} {\bibfnamefont
  {J.}~\bibnamefont {Ye}},\ }\href {https://doi.org/10.1038/nature20811}
  {\bibfield  {journal} {\bibinfo  {journal} {Nature}\ }\textbf {\bibinfo
  {volume} {542}},\ \bibinfo {pages} {66} (\bibinfo {year} {2017})}\BibitemShut
  {NoStop}%
\bibitem [{\citenamefont {Chalopin}\ \emph {et~al.}(2020)\citenamefont
  {Chalopin}, \citenamefont {Satoor}, \citenamefont {Evrard}, \citenamefont
  {Makhalor}, \citenamefont {Dalibard}, \citenamefont {Lopes},\ and\
  \citenamefont {Nascimbene}}]{cse20}%
  \BibitemOpen
  \bibfield  {author} {\bibinfo {author} {\bibfnamefont {T.}~\bibnamefont
  {Chalopin}}, \bibinfo {author} {\bibfnamefont {T.}~\bibnamefont {Satoor}},
  \bibinfo {author} {\bibfnamefont {A.}~\bibnamefont {Evrard}}, \bibinfo
  {author} {\bibfnamefont {V.}~\bibnamefont {Makhalor}}, \bibinfo {author}
  {\bibfnamefont {J.}~\bibnamefont {Dalibard}}, \bibinfo {author}
  {\bibfnamefont {R.}~\bibnamefont {Lopes}},\ and\ \bibinfo {author}
  {\bibfnamefont {S.}~\bibnamefont {Nascimbene}},\ }\href
  {https://doi.org/10.1038/s41567-020-0942-5} {\bibfield  {journal} {\bibinfo
  {journal} {Nat. Phys.}\ }\textbf {\bibinfo {volume} {16}},\ \bibinfo {pages}
  {1017} (\bibinfo {year} {2020})}\BibitemShut {NoStop}%
\bibitem [{\citenamefont {Kanungo}\ \emph {et~al.}(2022)\citenamefont
  {Kanungo}, \citenamefont {Whalen}, \citenamefont {Lu}, \citenamefont {Yuan},
  \citenamefont {Dasgupta}, \citenamefont {Dunning}, \citenamefont {Hazzard},\
  and\ \citenamefont {Killian}}]{kwl22}%
  \BibitemOpen
  \bibfield  {author} {\bibinfo {author} {\bibfnamefont {S.~K.}\ \bibnamefont
  {Kanungo}}, \bibinfo {author} {\bibfnamefont {J.~D.}\ \bibnamefont {Whalen}},
  \bibinfo {author} {\bibfnamefont {Y.}~\bibnamefont {Lu}}, \bibinfo {author}
  {\bibfnamefont {M.}~\bibnamefont {Yuan}}, \bibinfo {author} {\bibfnamefont
  {S.}~\bibnamefont {Dasgupta}}, \bibinfo {author} {\bibfnamefont {F.~B.}\
  \bibnamefont {Dunning}}, \bibinfo {author} {\bibfnamefont {K.~R.~A.}\
  \bibnamefont {Hazzard}},\ and\ \bibinfo {author} {\bibfnamefont {T.~C.}\
  \bibnamefont {Killian}},\ }\href {https://doi.org/10.1038/s41467-022-28550-y}
  {\bibfield  {journal} {\bibinfo  {journal} {Nat. Comm.}\ }\textbf {\bibinfo
  {volume} {13}},\ \bibinfo {pages} {972} (\bibinfo {year} {2022})}\BibitemShut
  {NoStop}%
\bibitem [{\citenamefont {Yoshida}\ \emph {et~al.}(2012)\citenamefont
  {Yoshida}, \citenamefont {Reinhold}, \citenamefont {Burgd\"orfer},
  \citenamefont {Ye},\ and\ \citenamefont {Dunning}}]{yrb12}%
  \BibitemOpen
  \bibfield  {author} {\bibinfo {author} {\bibfnamefont {S.}~\bibnamefont
  {Yoshida}}, \bibinfo {author} {\bibfnamefont {C.~O.}\ \bibnamefont
  {Reinhold}}, \bibinfo {author} {\bibfnamefont {J.}~\bibnamefont
  {Burgd\"orfer}}, \bibinfo {author} {\bibfnamefont {S.}~\bibnamefont {Ye}},\
  and\ \bibinfo {author} {\bibfnamefont {F.~B.}\ \bibnamefont {Dunning}},\
  }\href {https://doi.org/10.1103/PhysRevA.86.043415} {\bibfield  {journal}
  {\bibinfo  {journal} {Phys. Rev. A}\ }\textbf {\bibinfo {volume} {86}},\
  \bibinfo {pages} {043415} (\bibinfo {year} {2012})}\BibitemShut {NoStop}%
\bibitem [{\citenamefont {Oliver}\ \emph {et~al.}(2023)\citenamefont {Oliver},
  \citenamefont {Smith}, \citenamefont {Easton}, \citenamefont {Salerno},
  \citenamefont {Guarrera}, \citenamefont {Goldman}, \citenamefont
  {Barontini},\ and\ \citenamefont {Price}}]{ose23}%
  \BibitemOpen
  \bibfield  {author} {\bibinfo {author} {\bibfnamefont {C.}~\bibnamefont
  {Oliver}}, \bibinfo {author} {\bibfnamefont {A.}~\bibnamefont {Smith}},
  \bibinfo {author} {\bibfnamefont {T.}~\bibnamefont {Easton}}, \bibinfo
  {author} {\bibfnamefont {G.}~\bibnamefont {Salerno}}, \bibinfo {author}
  {\bibfnamefont {V.}~\bibnamefont {Guarrera}}, \bibinfo {author}
  {\bibfnamefont {N.}~\bibnamefont {Goldman}}, \bibinfo {author} {\bibfnamefont
  {G.}~\bibnamefont {Barontini}},\ and\ \bibinfo {author} {\bibfnamefont
  {H.~M.}\ \bibnamefont {Price}},\ }\href
  {https://doi.org/10.1103/PhysRevResearch.5.033001} {\bibfield  {journal}
  {\bibinfo  {journal} {Phys. Rev. Res.}\ }\textbf {\bibinfo {volume} {5}},\
  \bibinfo {pages} {033001} (\bibinfo {year} {2023})}\BibitemShut {NoStop}%
\bibitem [{\citenamefont {Meier}\ \emph {et~al.}(2016)\citenamefont {Meier},
  \citenamefont {An},\ and\ \citenamefont {Gadway}}]{mag16}%
  \BibitemOpen
  \bibfield  {author} {\bibinfo {author} {\bibfnamefont {E.}~\bibnamefont
  {Meier}}, \bibinfo {author} {\bibfnamefont {F.}~\bibnamefont {An}},\ and\
  \bibinfo {author} {\bibfnamefont {B.}~\bibnamefont {Gadway}},\ }\href
  {https://doi.org/10.1038/ncomms13986} {\bibfield  {journal} {\bibinfo
  {journal} {Nat. Comm.}\ }\textbf {\bibinfo {volume} {7}},\ \bibinfo {pages}
  {13986} (\bibinfo {year} {2016})}\BibitemShut {NoStop}%
\bibitem [{\citenamefont {Viebahn}\ \emph {et~al.}(2019)\citenamefont
  {Viebahn}, \citenamefont {Sbroscia}, \citenamefont {Carter}, \citenamefont
  {Yu},\ and\ \citenamefont {Schneider}}]{vsc19}%
  \BibitemOpen
  \bibfield  {author} {\bibinfo {author} {\bibfnamefont {K.}~\bibnamefont
  {Viebahn}}, \bibinfo {author} {\bibfnamefont {M.}~\bibnamefont {Sbroscia}},
  \bibinfo {author} {\bibfnamefont {E.}~\bibnamefont {Carter}}, \bibinfo
  {author} {\bibfnamefont {J.-C.}\ \bibnamefont {Yu}},\ and\ \bibinfo {author}
  {\bibfnamefont {U.}~\bibnamefont {Schneider}},\ }\href
  {https://doi.org/10.1103/PhysRevLett.122.110404} {\bibfield  {journal}
  {\bibinfo  {journal} {Phys. Rev. Lett.}\ }\textbf {\bibinfo {volume} {122}},\
  \bibinfo {pages} {110404} (\bibinfo {year} {2019})}\BibitemShut {NoStop}%
\bibitem [{\citenamefont {Kang}\ \emph {et~al.}(2020)\citenamefont {Kang},
  \citenamefont {Han},\ and\ \citenamefont {Shin}}]{khs20}%
  \BibitemOpen
  \bibfield  {author} {\bibinfo {author} {\bibfnamefont {J.~H.}\ \bibnamefont
  {Kang}}, \bibinfo {author} {\bibfnamefont {J.~H.}\ \bibnamefont {Han}},\ and\
  \bibinfo {author} {\bibfnamefont {Y.}~\bibnamefont {Shin}},\ }\href
  {https://doi.org/10.1088/1367-2630/ab61d7} {\bibfield  {journal} {\bibinfo
  {journal} {New J. Phys.}\ }\textbf {\bibinfo {volume} {22}},\ \bibinfo
  {pages} {013023} (\bibinfo {year} {2020})}\BibitemShut {NoStop}%
\bibitem [{\citenamefont {Cai}\ \emph {et~al.}(2019)\citenamefont {Cai},
  \citenamefont {Liu}, \citenamefont {Wu}, \citenamefont {He}, \citenamefont
  {Zhu}, \citenamefont {Zhang},\ and\ \citenamefont {Wang}}]{clw19}%
  \BibitemOpen
  \bibfield  {author} {\bibinfo {author} {\bibfnamefont {H.}~\bibnamefont
  {Cai}}, \bibinfo {author} {\bibfnamefont {J.}~\bibnamefont {Liu}}, \bibinfo
  {author} {\bibfnamefont {J.}~\bibnamefont {Wu}}, \bibinfo {author}
  {\bibfnamefont {Y.}~\bibnamefont {He}}, \bibinfo {author} {\bibfnamefont
  {S.-Y.}\ \bibnamefont {Zhu}}, \bibinfo {author} {\bibfnamefont {J.-X.}\
  \bibnamefont {Zhang}},\ and\ \bibinfo {author} {\bibfnamefont {D.-W.}\
  \bibnamefont {Wang}},\ }\href
  {https://doi.org/10.1103/PhysRevLett.122.023601} {\bibfield  {journal}
  {\bibinfo  {journal} {Phys. Rev. Lett.}\ }\textbf {\bibinfo {volume} {122}},\
  \bibinfo {pages} {023601} (\bibinfo {year} {2019})}\BibitemShut {NoStop}%
\bibitem [{\citenamefont {Meier}(2018)}]{mei18}%
  \BibitemOpen
  \bibfield  {author} {\bibinfo {author} {\bibfnamefont {E.~J.}\ \bibnamefont
  {Meier}},\ }\href {https://doi.org/10.1126/science.aat3406} {\bibfield
  {journal} {\bibinfo  {journal} {Science}\ }\textbf {\bibinfo {volume}
  {362}},\ \bibinfo {pages} {929} (\bibinfo {year} {2018})}\BibitemShut
  {NoStop}%
\bibitem [{\citenamefont {Price}\ \emph {et~al.}(2015)\citenamefont {Price},
  \citenamefont {Zilberberg}, \citenamefont {Ozawa}, \citenamefont
  {Carusotto},\ and\ \citenamefont {Goldman}}]{pzo18}%
  \BibitemOpen
  \bibfield  {author} {\bibinfo {author} {\bibfnamefont {H.~M.}\ \bibnamefont
  {Price}}, \bibinfo {author} {\bibfnamefont {O.}~\bibnamefont {Zilberberg}},
  \bibinfo {author} {\bibfnamefont {T.}~\bibnamefont {Ozawa}}, \bibinfo
  {author} {\bibfnamefont {I.}~\bibnamefont {Carusotto}},\ and\ \bibinfo
  {author} {\bibfnamefont {N.}~\bibnamefont {Goldman}},\ }\href
  {https://doi.org/10.1103/PhysRevLett.115.195303} {\bibfield  {journal}
  {\bibinfo  {journal} {Phys. Rev. Lett.}\ }\textbf {\bibinfo {volume} {115}},\
  \bibinfo {pages} {195303} (\bibinfo {year} {2015})}\BibitemShut {NoStop}%
\bibitem [{\citenamefont {Chen}\ \emph {et~al.}(2023)\citenamefont {Chen},
  \citenamefont {Huang}, \citenamefont {Velkovsky}, \citenamefont {Hazzard},
  \citenamefont {Covey},\ and\ \citenamefont {Gadway}}]{chv23}%
  \BibitemOpen
  \bibfield  {author} {\bibinfo {author} {\bibfnamefont {T.}~\bibnamefont
  {Chen}}, \bibinfo {author} {\bibfnamefont {C.}~\bibnamefont {Huang}},
  \bibinfo {author} {\bibfnamefont {I.}~\bibnamefont {Velkovsky}}, \bibinfo
  {author} {\bibfnamefont {K.~R.~A.}\ \bibnamefont {Hazzard}}, \bibinfo
  {author} {\bibfnamefont {J.~P.}\ \bibnamefont {Covey}},\ and\ \bibinfo
  {author} {\bibfnamefont {B.}~\bibnamefont {Gadway}},\ }\href@noop {} {}
  (\bibinfo {year} {2023}),\ \Eprint {https://arxiv.org/abs/2306.00883}
  {arXiv:2306.00883 [cond-mat.quant-gas]} \BibitemShut {NoStop}%
\bibitem [{\citenamefont {Su}\ \emph {et~al.}(1979)\citenamefont {Su},
  \citenamefont {Schrieffer},\ and\ \citenamefont {Heeger}}]{ssh79}%
  \BibitemOpen
  \bibfield  {author} {\bibinfo {author} {\bibfnamefont {W.~P.}\ \bibnamefont
  {Su}}, \bibinfo {author} {\bibfnamefont {J.~R.}\ \bibnamefont {Schrieffer}},\
  and\ \bibinfo {author} {\bibfnamefont {A.~J.}\ \bibnamefont {Heeger}},\
  }\href {https://doi.org/10.1103/PhysRevLett.42.1698} {\bibfield  {journal}
  {\bibinfo  {journal} {Phys. Rev. Lett.}\ }\textbf {\bibinfo {volume} {42}},\
  \bibinfo {pages} {1698} (\bibinfo {year} {1979})}\BibitemShut {NoStop}%
\bibitem [{\citenamefont {Chiu}\ \emph {et~al.}(2016)\citenamefont {Chiu},
  \citenamefont {Teo}, \citenamefont {Schnyder},\ and\ \citenamefont
  {Ryu}}]{cts16}%
  \BibitemOpen
  \bibfield  {author} {\bibinfo {author} {\bibfnamefont {C.-K.}\ \bibnamefont
  {Chiu}}, \bibinfo {author} {\bibfnamefont {J.~C.~Y.}\ \bibnamefont {Teo}},
  \bibinfo {author} {\bibfnamefont {A.~P.}\ \bibnamefont {Schnyder}},\ and\
  \bibinfo {author} {\bibfnamefont {S.}~\bibnamefont {Ryu}},\ }\href
  {https://doi.org/10.1103/RevModPhys.88.035005} {\bibfield  {journal}
  {\bibinfo  {journal} {Rev. Mod. Phys.}\ }\textbf {\bibinfo {volume} {88}},\
  \bibinfo {pages} {035005} (\bibinfo {year} {2016})}\BibitemShut {NoStop}%
\bibitem [{\citenamefont {Atala}\ \emph {et~al.}(2013)\citenamefont {Atala},
  \citenamefont {Aidelsburger}, \citenamefont {Barreiro}, \citenamefont
  {Abania}, \citenamefont {Kitagawa}, \citenamefont {Demler},\ and\
  \citenamefont {Bloch}}]{aab13}%
  \BibitemOpen
  \bibfield  {author} {\bibinfo {author} {\bibfnamefont {M.}~\bibnamefont
  {Atala}}, \bibinfo {author} {\bibfnamefont {M.}~\bibnamefont {Aidelsburger}},
  \bibinfo {author} {\bibfnamefont {J.~J.}\ \bibnamefont {Barreiro}}, \bibinfo
  {author} {\bibfnamefont {D.}~\bibnamefont {Abania}}, \bibinfo {author}
  {\bibfnamefont {T.}~\bibnamefont {Kitagawa}}, \bibinfo {author}
  {\bibfnamefont {E.}~\bibnamefont {Demler}},\ and\ \bibinfo {author}
  {\bibfnamefont {I.}~\bibnamefont {Bloch}},\ }\href@noop {} {\bibfield
  {journal} {\bibinfo  {journal} {Nature Physics}\ }\textbf {\bibinfo {volume}
  {9}},\ \bibinfo {pages} {795} (\bibinfo {year} {2013})}\BibitemShut {NoStop}%
\bibitem [{\citenamefont {St-Jean}\ \emph {et~al.}(2017)\citenamefont
  {St-Jean}, \citenamefont {Godblot}, \citenamefont {Galopin}, \citenamefont
  {Lemaitre}, \citenamefont {Ozawa}, \citenamefont {LeGratiet}, \citenamefont
  {Sagnes}, \citenamefont {Bloch},\ and\ \citenamefont {Amo}}]{sgg17}%
  \BibitemOpen
  \bibfield  {author} {\bibinfo {author} {\bibfnamefont {P.}~\bibnamefont
  {St-Jean}}, \bibinfo {author} {\bibfnamefont {V.}~\bibnamefont {Godblot}},
  \bibinfo {author} {\bibfnamefont {E.}~\bibnamefont {Galopin}}, \bibinfo
  {author} {\bibfnamefont {A.}~\bibnamefont {Lemaitre}}, \bibinfo {author}
  {\bibfnamefont {T.}~\bibnamefont {Ozawa}}, \bibinfo {author} {\bibfnamefont
  {L.}~\bibnamefont {LeGratiet}}, \bibinfo {author} {\bibfnamefont
  {I.}~\bibnamefont {Sagnes}}, \bibinfo {author} {\bibfnamefont
  {J.}~\bibnamefont {Bloch}},\ and\ \bibinfo {author} {\bibfnamefont
  {A.}~\bibnamefont {Amo}},\ }\href {https://doi.org/10.1038/s41566-017-0006-2}
  {\bibfield  {journal} {\bibinfo  {journal} {Nature Photonics}\ }\textbf
  {\bibinfo {volume} {11}},\ \bibinfo {pages} {651} (\bibinfo {year}
  {2017})}\BibitemShut {NoStop}%
\bibitem [{\citenamefont {Léséleuc}\ \emph {et~al.}(2019)\citenamefont
  {Léséleuc}, \citenamefont {Lienhard}, \citenamefont {Scholl}, \citenamefont
  {Barredo}, \citenamefont {Weber}, \citenamefont {Lang}, \citenamefont
  {Büchler}, \citenamefont {Lahaye},\ and\ \citenamefont {Browaeys}}]{llp19}%
  \BibitemOpen
  \bibfield  {author} {\bibinfo {author} {\bibfnamefont {S.}~\bibnamefont
  {Léséleuc}}, \bibinfo {author} {\bibfnamefont {V.}~\bibnamefont
  {Lienhard}}, \bibinfo {author} {\bibfnamefont {P.}~\bibnamefont {Scholl}},
  \bibinfo {author} {\bibfnamefont {D.}~\bibnamefont {Barredo}}, \bibinfo
  {author} {\bibfnamefont {S.}~\bibnamefont {Weber}}, \bibinfo {author}
  {\bibfnamefont {N.}~\bibnamefont {Lang}}, \bibinfo {author} {\bibfnamefont
  {H.~P.}\ \bibnamefont {Büchler}}, \bibinfo {author} {\bibfnamefont
  {T.}~\bibnamefont {Lahaye}},\ and\ \bibinfo {author} {\bibfnamefont
  {A.}~\bibnamefont {Browaeys}},\ }\href
  {https://doi.org/10.1126/science.aav9105} {\bibfield  {journal} {\bibinfo
  {journal} {Science}\ }\textbf {\bibinfo {volume} {365}},\ \bibinfo {pages}
  {775} (\bibinfo {year} {2019})}\BibitemShut {NoStop}%
\bibitem [{\citenamefont {Hazzard}\ and\ \citenamefont
  {Gadway}(2023)}]{hga23a}%
  \BibitemOpen
  \bibfield  {author} {\bibinfo {author} {\bibfnamefont {K.~R.~A.}\
  \bibnamefont {Hazzard}}\ and\ \bibinfo {author} {\bibfnamefont
  {B.}~\bibnamefont {Gadway}},\ }\href {https://doi.org/10.1063/PT.3.5225}
  {\bibfield  {journal} {\bibinfo  {journal} {Physics Today}\ }\textbf
  {\bibinfo {volume} {76}},\ \bibinfo {pages} {62} (\bibinfo {year}
  {2023})}\BibitemShut {NoStop}%
\bibitem [{\citenamefont {Stellmer}\ \emph {et~al.}(2013)\citenamefont
  {Stellmer}, \citenamefont {Schreck},\ and\ \citenamefont {Killian}}]{ssk13}%
  \BibitemOpen
  \bibfield  {author} {\bibinfo {author} {\bibfnamefont {S.}~\bibnamefont
  {Stellmer}}, \bibinfo {author} {\bibfnamefont {F.}~\bibnamefont {Schreck}},\
  and\ \bibinfo {author} {\bibfnamefont {T.~C.}\ \bibnamefont {Killian}},\
  }\href@noop {} {\emph {\bibinfo {title} {Annual Review of Cold Atoms and
  Molecules}}},\ Vol.~\bibinfo {volume} {2}\ (\bibinfo  {publisher} {World
  Scientific},\ \bibinfo {address} {Singapore},\ \bibinfo {year} {2013})\ pp.\
  \bibinfo {pages} {1--80}\BibitemShut {NoStop}%
\bibitem [{\citenamefont {de~Escobar}\ \emph {et~al.}(2009)\citenamefont
  {de~Escobar}, \citenamefont {Mickelson}, \citenamefont {Yan}, \citenamefont
  {DeSalvo}, \citenamefont {Nagel},\ and\ \citenamefont {Killian}}]{emm09}%
  \BibitemOpen
  \bibfield  {author} {\bibinfo {author} {\bibfnamefont {Y.~N.~M.}\
  \bibnamefont {de~Escobar}}, \bibinfo {author} {\bibfnamefont {P.~G.}\
  \bibnamefont {Mickelson}}, \bibinfo {author} {\bibfnamefont {M.}~\bibnamefont
  {Yan}}, \bibinfo {author} {\bibfnamefont {B.~J.}\ \bibnamefont {DeSalvo}},
  \bibinfo {author} {\bibfnamefont {S.~B.}\ \bibnamefont {Nagel}},\ and\
  \bibinfo {author} {\bibfnamefont {T.~C.}\ \bibnamefont {Killian}},\ }\href
  {https://doi.org/10.1103/PhysRevLett.103.200402} {\bibfield  {journal}
  {\bibinfo  {journal} {Phys. Rev. Lett.}\ }\textbf {\bibinfo {volume} {103}},\
  \bibinfo {pages} {200402} (\bibinfo {year} {2009})}\BibitemShut {NoStop}%
\bibitem [{\citenamefont {Ding}\ \emph {et~al.}(2018)\citenamefont {Ding},
  \citenamefont {Whalen}, \citenamefont {Kanungo}, \citenamefont {Killian},
  \citenamefont {Dunning}, \citenamefont {Yoshida},\ and\ \citenamefont
  {Burgdörfer}}]{dwk18}%
  \BibitemOpen
  \bibfield  {author} {\bibinfo {author} {\bibfnamefont {R.}~\bibnamefont
  {Ding}}, \bibinfo {author} {\bibfnamefont {J.~D.}\ \bibnamefont {Whalen}},
  \bibinfo {author} {\bibfnamefont {S.~K.}\ \bibnamefont {Kanungo}}, \bibinfo
  {author} {\bibfnamefont {T.~C.}\ \bibnamefont {Killian}}, \bibinfo {author}
  {\bibfnamefont {F.~B.}\ \bibnamefont {Dunning}}, \bibinfo {author}
  {\bibfnamefont {S.}~\bibnamefont {Yoshida}},\ and\ \bibinfo {author}
  {\bibfnamefont {J.}~\bibnamefont {Burgdörfer}},\ }\href
  {https://doi.org/10.1103/physreva.98.042505} {\bibfield  {journal} {\bibinfo
  {journal} {Physical Review A}\ }\textbf {\bibinfo {volume} {98}},\ \bibinfo
  {pages} {042505} (\bibinfo {year} {2018})}\BibitemShut {NoStop}%
\bibitem [{\citenamefont {Asb{\'o}th}\ \emph {et~al.}(2016)\citenamefont
  {Asb{\'o}th}, \citenamefont {Oroszl{\'a}ny},\ and\ \citenamefont
  {P{\'a}lyi}}]{aop16}%
  \BibitemOpen
  \bibfield  {author} {\bibinfo {author} {\bibfnamefont {J.~K.}\ \bibnamefont
  {Asb{\'o}th}}, \bibinfo {author} {\bibfnamefont {L.}~\bibnamefont
  {Oroszl{\'a}ny}},\ and\ \bibinfo {author} {\bibfnamefont {A.}~\bibnamefont
  {P{\'a}lyi}},\ }\bibinfo {title} {The {Su-Schrieffer-Heeger (SSH)} model},\
  in\ \href {https://doi.org/10.1007/978-3-319-25607-8_1} {\emph {\bibinfo
  {booktitle} {A Short Course on Topological Insulators: Band Structure and
  Edge States in One and Two Dimensions}}}\ (\bibinfo  {publisher} {Springer
  International Publishing},\ \bibinfo {address} {Cham},\ \bibinfo {year}
  {2016})\ pp.\ \bibinfo {pages} {1--22}\BibitemShut {NoStop}%
\bibitem [{\citenamefont {Jackiw}(2007)}]{jac07}%
  \BibitemOpen
  \bibfield  {author} {\bibinfo {author} {\bibfnamefont {R.}~\bibnamefont
  {Jackiw}},\ }\href {https://doi.org/10.1063/1.2803825} {\bibfield  {journal}
  {\bibinfo  {journal} {AIP Conference Proceedings}\ }\textbf {\bibinfo
  {volume} {939}},\ \bibinfo {pages} {341} (\bibinfo {year}
  {2007})}\BibitemShut {NoStop}%
\bibitem [{\citenamefont {Tsomokos}\ \emph {et~al.}(2010)\citenamefont
  {Tsomokos}, \citenamefont {Ashhab},\ and\ \citenamefont {Nori}}]{tso10}%
  \BibitemOpen
  \bibfield  {author} {\bibinfo {author} {\bibfnamefont {D.~I.}\ \bibnamefont
  {Tsomokos}}, \bibinfo {author} {\bibfnamefont {S.}~\bibnamefont {Ashhab}},\
  and\ \bibinfo {author} {\bibfnamefont {F.}~\bibnamefont {Nori}},\ }\href
  {https://doi.org/10.1103/PhysRevA.82.052311} {\bibfield  {journal} {\bibinfo
  {journal} {Phys. Rev. A}\ }\textbf {\bibinfo {volume} {82}},\ \bibinfo
  {pages} {052311} (\bibinfo {year} {2010})}\BibitemShut {NoStop}%
\bibitem [{\citenamefont {Juki\ifmmode~\acute{c}\else \'{c}\fi{}}\ and\
  \citenamefont {Buljan}(2013)}]{jbu13}%
  \BibitemOpen
  \bibfield  {author} {\bibinfo {author} {\bibfnamefont {D.}~\bibnamefont
  {Juki\ifmmode~\acute{c}\else \'{c}\fi{}}}\ and\ \bibinfo {author}
  {\bibfnamefont {H.}~\bibnamefont {Buljan}},\ }\href
  {https://doi.org/10.1103/PhysRevA.87.013814} {\bibfield  {journal} {\bibinfo
  {journal} {Phys. Rev. A}\ }\textbf {\bibinfo {volume} {87}},\ \bibinfo
  {pages} {013814} (\bibinfo {year} {2013})}\BibitemShut {NoStop}%
\bibitem [{\citenamefont {Boada}\ \emph {et~al.}(2015)\citenamefont {Boada},
  \citenamefont {Celi}, \citenamefont {Rodriguez-Laguna}, \citenamefont
  {Latorre},\ and\ \citenamefont {Lewenstein}}]{bcr15}%
  \BibitemOpen
  \bibfield  {author} {\bibinfo {author} {\bibfnamefont {O.}~\bibnamefont
  {Boada}}, \bibinfo {author} {\bibfnamefont {A.}~\bibnamefont {Celi}},
  \bibinfo {author} {\bibfnamefont {J.}~\bibnamefont {Rodriguez-Laguna}},
  \bibinfo {author} {\bibfnamefont {J.~I.}\ \bibnamefont {Latorre}},\ and\
  \bibinfo {author} {\bibfnamefont {M.}~\bibnamefont {Lewenstein}},\
  }\href@noop {} {\bibfield  {journal} {\bibinfo  {journal} {New Journal of
  Physics}\ }\textbf {\bibinfo {volume} {17}},\ \bibinfo {pages} {045007}
  (\bibinfo {year} {2015})}\BibitemShut {NoStop}%
\bibitem [{\citenamefont {Bansil}\ \emph {et~al.}(2016)\citenamefont {Bansil},
  \citenamefont {Lin},\ and\ \citenamefont {Das}}]{bld16}%
  \BibitemOpen
  \bibfield  {author} {\bibinfo {author} {\bibfnamefont {A.}~\bibnamefont
  {Bansil}}, \bibinfo {author} {\bibfnamefont {H.}~\bibnamefont {Lin}},\ and\
  \bibinfo {author} {\bibfnamefont {T.}~\bibnamefont {Das}},\ }\href
  {https://doi.org/10.1103/RevModPhys.88.021004} {\bibfield  {journal}
  {\bibinfo  {journal} {Rev. Mod. Phys.}\ }\textbf {\bibinfo {volume} {88}},\
  \bibinfo {pages} {021004} (\bibinfo {year} {2016})}\BibitemShut {NoStop}%
\bibitem [{\citenamefont {Qi}\ and\ \citenamefont {Zhang}(2011)}]{qzh11}%
  \BibitemOpen
  \bibfield  {author} {\bibinfo {author} {\bibfnamefont {X.-L.}\ \bibnamefont
  {Qi}}\ and\ \bibinfo {author} {\bibfnamefont {S.-C.}\ \bibnamefont {Zhang}},\
  }\href {https://doi.org/10.1103/RevModPhys.83.1057} {\bibfield  {journal}
  {\bibinfo  {journal} {Rev. Mod. Phys.}\ }\textbf {\bibinfo {volume} {83}},\
  \bibinfo {pages} {1057} (\bibinfo {year} {2011})}\BibitemShut {NoStop}%
\bibitem [{\citenamefont {Bernevig}\ and\ \citenamefont
  {Hughes}(2017)}]{bhu17}%
  \BibitemOpen
  \bibfield  {author} {\bibinfo {author} {\bibfnamefont {B.~A.}\ \bibnamefont
  {Bernevig}}\ and\ \bibinfo {author} {\bibfnamefont {T.~L.}\ \bibnamefont
  {Hughes}},\ }\href@noop {} {\bibfield  {journal} {\bibinfo  {journal}
  {Science}\ }\textbf {\bibinfo {volume} {357}},\ \bibinfo {pages} {61}
  (\bibinfo {year} {2017})}\BibitemShut {NoStop}%
\bibitem [{\citenamefont {Roy}\ and\ \citenamefont {Harper}(2017)}]{rha17}%
  \BibitemOpen
  \bibfield  {author} {\bibinfo {author} {\bibfnamefont {R.}~\bibnamefont
  {Roy}}\ and\ \bibinfo {author} {\bibfnamefont {F.}~\bibnamefont {Harper}},\
  }\href {https://doi.org/10.1103/PhysRevB.96.155118} {\bibfield  {journal}
  {\bibinfo  {journal} {Phys. Rev. B}\ }\textbf {\bibinfo {volume} {96}},\
  \bibinfo {pages} {155118} (\bibinfo {year} {2017})}\BibitemShut {NoStop}%
\bibitem [{\citenamefont {McGinley}\ and\ \citenamefont
  {Cooper}(2019)}]{mco19}%
  \BibitemOpen
  \bibfield  {author} {\bibinfo {author} {\bibfnamefont {M.}~\bibnamefont
  {McGinley}}\ and\ \bibinfo {author} {\bibfnamefont {N.~R.}\ \bibnamefont
  {Cooper}},\ }\href {https://doi.org/10.1103/PhysRevB.99.075148} {\bibfield
  {journal} {\bibinfo  {journal} {Phys. Rev. B}\ }\textbf {\bibinfo {volume}
  {99}},\ \bibinfo {pages} {075148} (\bibinfo {year} {2019})}\BibitemShut
  {NoStop}%
\bibitem [{\citenamefont {Sundar}\ \emph {et~al.}(2018)\citenamefont {Sundar},
  \citenamefont {Gadway},\ and\ \citenamefont {Hazzard}}]{sgh18}%
  \BibitemOpen
  \bibfield  {author} {\bibinfo {author} {\bibfnamefont {B.}~\bibnamefont
  {Sundar}}, \bibinfo {author} {\bibfnamefont {B.}~\bibnamefont {Gadway}},\
  and\ \bibinfo {author} {\bibfnamefont {K.~R.~A.}\ \bibnamefont {Hazzard}},\
  }\href {https://doi.org/10.1038/s41598-018-21699-x} {\bibfield  {journal}
  {\bibinfo  {journal} {Scientific Reports}\ }\textbf {\bibinfo {volume} {8}},\
  \bibinfo {pages} {3422} (\bibinfo {year} {2018})}\BibitemShut {NoStop}%
\bibitem [{\citenamefont {Sundar}\ \emph {et~al.}(2019)\citenamefont {Sundar},
  \citenamefont {Thibodeau}, \citenamefont {Wang}, \citenamefont {Gadway},\
  and\ \citenamefont {Hazzard}}]{stw19}%
  \BibitemOpen
  \bibfield  {author} {\bibinfo {author} {\bibfnamefont {B.}~\bibnamefont
  {Sundar}}, \bibinfo {author} {\bibfnamefont {M.}~\bibnamefont {Thibodeau}},
  \bibinfo {author} {\bibfnamefont {Z.}~\bibnamefont {Wang}}, \bibinfo {author}
  {\bibfnamefont {B.}~\bibnamefont {Gadway}},\ and\ \bibinfo {author}
  {\bibfnamefont {K.~R.~A.}\ \bibnamefont {Hazzard}},\ }\href
  {https://doi.org/10.1103/PhysRevA.99.013624} {\bibfield  {journal} {\bibinfo
  {journal} {Phys. Rev. A}\ }\textbf {\bibinfo {volume} {99}},\ \bibinfo
  {pages} {013624} (\bibinfo {year} {2019})}\BibitemShut {NoStop}%
\bibitem [{\citenamefont {Andersen}(2022)}]{and22}%
  \BibitemOpen
  \bibfield  {author} {\bibinfo {author} {\bibfnamefont {M.~F.}\ \bibnamefont
  {Andersen}},\ }\href {https://doi.org/10.1080/23746149.2022.2064231}
  {\bibfield  {journal} {\bibinfo  {journal} {Advances in Physics: X}\ }\textbf
  {\bibinfo {volume} {7}},\ \bibinfo {pages} {2064231} (\bibinfo {year}
  {2022})}\BibitemShut {NoStop}%
\bibitem [{\citenamefont {Browaeys}\ and\ \citenamefont
  {Lahaye}(2020)}]{bla20}%
  \BibitemOpen
  \bibfield  {author} {\bibinfo {author} {\bibfnamefont {A.}~\bibnamefont
  {Browaeys}}\ and\ \bibinfo {author} {\bibfnamefont {T.}~\bibnamefont
  {Lahaye}},\ }\href {https://doi.org/10.1038/s41567-019-0733-z} {\bibfield
  {journal} {\bibinfo  {journal} {Nature Physics}\ }\textbf {\bibinfo {volume}
  {16}},\ \bibinfo {pages} {132} (\bibinfo {year} {2020})}\BibitemShut
  {NoStop}%
\bibitem [{\citenamefont {Kauffman}\ and\ \citenamefont {Ni}(2021)}]{kni21}%
  \BibitemOpen
  \bibfield  {author} {\bibinfo {author} {\bibfnamefont {A.~M.}\ \bibnamefont
  {Kauffman}}\ and\ \bibinfo {author} {\bibfnamefont {K.~K.}\ \bibnamefont
  {Ni}},\ }\href {https://doi.org/10.1038/s41567-021-01357-2} {\bibfield
  {journal} {\bibinfo  {journal} {Nature Physics}\ }\textbf {\bibinfo {volume}
  {17}},\ \bibinfo {pages} {1324} (\bibinfo {year} {2021})}\BibitemShut
  {NoStop}%
\bibitem [{\citenamefont {Ebadi}\ \emph {et~al.}(2021)\citenamefont {Ebadi},
  \citenamefont {Wang}, \citenamefont {Levine}, \citenamefont {Keesling},
  \citenamefont {Semeghini},\ and\ \citenamefont {Omran}}]{ewl21}%
  \BibitemOpen
  \bibfield  {author} {\bibinfo {author} {\bibfnamefont {S.}~\bibnamefont
  {Ebadi}}, \bibinfo {author} {\bibfnamefont {T.~T.}\ \bibnamefont {Wang}},
  \bibinfo {author} {\bibfnamefont {H.}~\bibnamefont {Levine}}, \bibinfo
  {author} {\bibfnamefont {A.}~\bibnamefont {Keesling}}, \bibinfo {author}
  {\bibfnamefont {G.}~\bibnamefont {Semeghini}},\ and\ \bibinfo {author}
  {\bibfnamefont {A.}~\bibnamefont {Omran}},\ }\href
  {https://doi.org/10.1038/s41586-021-03582-4} {\bibfield  {journal} {\bibinfo
  {journal} {Nature}\ }\textbf {\bibinfo {volume} {595}},\ \bibinfo {pages}
  {227} (\bibinfo {year} {2021})}\BibitemShut {NoStop}%
\bibitem [{\citenamefont {Scholl}\ \emph {et~al.}(2021)\citenamefont {Scholl},
  \citenamefont {Schuler}, \citenamefont {Williams}, \citenamefont
  {Eberharter}, \citenamefont {Barredo}, \citenamefont {Schmik}, \citenamefont
  {Lienhard}, \citenamefont {Henry}, \citenamefont {Lang}, \citenamefont
  {Lahaye}, \citenamefont {Lauchli},\ and\ \citenamefont {Browaeys}}]{ssw21}%
  \BibitemOpen
  \bibfield  {author} {\bibinfo {author} {\bibfnamefont {P.}~\bibnamefont
  {Scholl}}, \bibinfo {author} {\bibfnamefont {M.}~\bibnamefont {Schuler}},
  \bibinfo {author} {\bibfnamefont {H.~J.}\ \bibnamefont {Williams}}, \bibinfo
  {author} {\bibfnamefont {A.}~\bibnamefont {Eberharter}}, \bibinfo {author}
  {\bibfnamefont {D.}~\bibnamefont {Barredo}}, \bibinfo {author} {\bibfnamefont
  {K.~N.}\ \bibnamefont {Schmik}}, \bibinfo {author} {\bibfnamefont
  {V.}~\bibnamefont {Lienhard}}, \bibinfo {author} {\bibfnamefont {L.~P.}\
  \bibnamefont {Henry}}, \bibinfo {author} {\bibfnamefont {T.~C.}\ \bibnamefont
  {Lang}}, \bibinfo {author} {\bibfnamefont {T.}~\bibnamefont {Lahaye}},
  \bibinfo {author} {\bibfnamefont {A.~M.}\ \bibnamefont {Lauchli}},\ and\
  \bibinfo {author} {\bibfnamefont {A.}~\bibnamefont {Browaeys}},\ }\href@noop
  {} {\bibfield  {journal} {\bibinfo  {journal} {Nature}\ }\textbf {\bibinfo
  {volume} {595}},\ \bibinfo {pages} {233} (\bibinfo {year}
  {2021})}\BibitemShut {NoStop}%
\end{thebibliography}
\end{document}